\documentclass[twocolumn,superscriptaddress,showkeys]{revtex4}
\usepackage{amssymb,epsf,epsfig,amsmath,color}

\newcommand{\lsim}{
\mathrel{\hbox{\rlap{\hbox{\lower4pt\hbox{$\sim$}}}\hbox{$<$}}}}
\newcommand{\gsim}{
\mathrel{\hbox{\rlap{\hbox{\lower4pt\hbox{$\sim$}}}\hbox{$>$}}}}

\def\D0{D\O }

\begin{document}
\begin{titlepage}
\vspace*{1.7truecm}
\begin{flushright}
Nikhef-2010-010
\end{flushright}

\vspace{1.6truecm}

\begin{center}
\boldmath
{\Large{\bf New Strategy for $B_s$ Branching Ratio Measurements

\vspace*{0.2truecm}

and the Search for New Physics in $B^0_s\to \mu^+\mu^-$}}
\unboldmath
\end{center}

\vspace{1.2truecm}

\begin{center}
{\bf Robert Fleischer, Nicola Serra and Niels Tuning}

\vspace{0.5truecm}

{\sl Nikhef, Science Park 105, NL-1098 XG Amsterdam, The Netherlands}

\end{center}

\vspace*{1.7cm}

\begin{center}
\large{\bf Abstract}\\

\vspace*{0.6truecm}

\begin{tabular}{p{14.5truecm}}
The LHCb experiment at CERN's Large Hadron Collider will soon allow us to enter
a new era in the exploration of $B_s$ decays. A particularly promising channel for 
the search of ``new physics" is $B^0_s\to\mu^+\mu^-$. The systematic key uncertainty 
affecting the measurement of this -- and in fact all $B_s$-decay branching ratios -- is 
the ratio of fragmentation functions $f_d/f_s$. As the presently available methods for 
determining $f_d/f_s$ are not sufficient to meet the high precision at LHCb, we propose 
a new strategy using $\bar B^0_s\to D_s^+\pi^-$ and $\bar B^0_d\to D^+K^-$. It allows 
us to obtain a lower experimental bound on $\mbox{BR}(B^0_s\to\mu^+\mu^-)$ which 
offers a powerful probe for new physics. In order to go beyond this bound and to 
determine $f_d/f_s$ with a theoretical precision matching the experimental one it is 
sufficient to know the $SU(3)$-breaking correction to a form-factor ratio from
nonperturbative methods at the level of 20\%. Thanks to our strategy, we can detect 
new physics in $B^0_s\to\mu^+\mu^-$ at LHCb with $5 \sigma$ for a branching ratio 
as small as twice the Standard-Model value, which represents an improvement of 
the new-physics reach by about a factor of 2 with respect to the current LHCb 
expectation.
\end{tabular}

\end{center}

\vspace*{1.7truecm}

\vfill

\noindent
April 2010

\end{titlepage}

\newpage
\thispagestyle{empty}
\mbox{}

\newpage
\thispagestyle{empty}
\mbox{}

\rule{0cm}{23cm}

\newpage
\thispagestyle{empty}
\mbox{}

\setcounter{page}{0}

\preprint{Nikhef-2010-010}

\date{April 22, 2010; published 31 August 2010}

\title{\boldmath New Strategy for $B_s$ Branching Ratio Measurements\\
and the Search for New Physics in $B^0_s\to \mu^+\mu^-$\unboldmath}

\author{Robert Fleischer}
\affiliation{Nikhef, Science Park 105, NL-1098 XG Amsterdam, The Netherlands}

\author{Nicola Serra}
\affiliation{Nikhef, Science Park 105, NL-1098 XG Amsterdam, The Netherlands}

\author{Niels Tuning}
\affiliation{Nikhef, Science Park 105, NL-1098 XG Amsterdam, The Netherlands}

\begin{abstract}
\vspace{0.2cm}\noindent
The LHCb experiment at CERN's Large Hadron Collider will soon allow us to enter
a new era in the exploration of $B_s$ decays. A particularly promising channel for 
the search of ``new physics" is $B^0_s\to\mu^+\mu^-$. The systematic key uncertainty 
affecting the measurement of this -- and in fact all $B_s$-decay branching ratios -- is 
the ratio of fragmentation functions $f_d/f_s$. As the presently available methods for 
determining $f_d/f_s$ are not sufficient to meet the high precision at LHCb, we propose 
a new strategy using $\bar B^0_s\to D_s^+\pi^-$ and $\bar B^0_d\to D^+K^-$. It allows 
us to obtain a lower experimental bound on $\mbox{BR}(B^0_s\to\mu^+\mu^-)$ which 
offers a powerful probe for new physics. In order to go beyond this bound and to 
determine $f_d/f_s$ with a theoretical precision matching the experimental one it is 
sufficient to know the $SU(3)$-breaking correction to a form-factor ratio from
nonperturbative methods at the level of 20\%. Thanks to our strategy, we can detect 
new physics in $B^0_s\to\mu^+\mu^-$ at LHCb with $5 \sigma$ for a branching ratio 
as small as twice the Standard-Model value, which represents an improvement of 
the new-physics reach by about a factor of 2 with respect to the current LHCb 
expectation.
\end{abstract}


\maketitle

\section{Introduction}
In this decade, we will enter a new round in the precision testing of the
flavor sector of the Standard Model (SM) through $B$-meson decays. 
Currently the LHCb experiment at CERN's Large Hardon Collider (LHC) 
is starting its first physics run. After pioneering results on the $B_s$ 
system by the CDF and \D0 collaborations at the Tevatron, LHCb will allow 
us to explore this still largely unexplored territory of the flavor-physics landscape
\cite{bib:LHCbRoadMap}.

In this respect, one of the most promising channels for detecting signals of 
``new physics'' (NP) is the rare decay $B^0_s\to\mu^+\mu^-$, which originates 
in the SM from ``penguin'' and box topologies, i.e.\ quantum loop processes.
The corresponding branching ratio is predicted as follows \cite{buras}:
\begin{equation}\label{Bsmumu-SM}
\mbox{BR}(B^0_s\to\mu^+\mu^-)|_{\rm SM}=(3.6\pm0.4)\times 10^{-9},
\end{equation}
where the error is fully dominated by a nonperturbative 
``bag parameter'' coming from lattice QCD. As is well known, this observable 
may be significantly enhanced through NP (for a review, see Ref.~\cite{buras}). 
The present upper bounds from the CDF and \D0 collaborations are 
still about 1 order of magnitude away from (\ref{Bsmumu-SM}) and read as 
$4.3\times 10^{-8}$ \cite{CDFbound} and $5.3\times 10^{-8}$ (95\% C.L.) 
\cite{D0bound}, respectively.
 
At LHCb, the extraction of $\mbox{BR}( B^0_s\to\mu^+\mu^-)$ will rely on 
normalization channels such as $B_u^+\to J/\psi K^+$, $B^0_d\to K^+\pi^-$ 
and/or $B_d^0 \to J/\psi K^{*0}$ in the following way:
\begin{equation}
\mbox{BR}(B^0_s\to\mu^+\mu^-)
=\mbox{BR}(B_q\to X)\frac{f_q}{f_s}
\frac{\epsilon_{X}}{\epsilon_{\mu\mu}}
\frac{N_{\mu\mu}}{N_{X}}, \label{BRmumu-exp}
\end{equation}
where the $\epsilon$ factors are total detector efficiencies and the $N$ factors denote 
the observed numbers of events. The $f_q$ are fragmentation functions, which describe 
the probability that a $b$ quark will fragment in a $\bar B_q$ meson ($q\in\{u,d,s\}$). In 
(\ref{BRmumu-exp}), $f_q/f_s$ is actually the major source of the systematic uncertainty, 
thereby limiting the ability to detect a $5 \sigma$ deviation from the SM at LHCb to 
$\mbox{BR}( B^0_s\to\mu^+\mu^-) > 11 \times 10^{-9}$ \cite{bib:LHCbRoadMap}. 

In this estimate, the current experimental knowledge of $f_d/f_s$ was assumed, which we 
summarize in Section~\ref{sec:status}. In Section~\ref{sec:strat}, we propose a new strategy 
to measure $f_d/f_s$ at LHCb. Its experimental prospects and theoretical limitations are discussed
in Sections~\ref{sec:exp} and \ref{sec:theo}, respectively. In Section~\ref{sec:impl}, we discuss
the implications for the search for NP with $B^0_s\to \mu^+\mu^-$ branching ratio, while we summarize 
our conclusions in Section~\ref{sec:concl}.

\section{Experimental Status of $f_d/f_s$}\label{sec:status}
The CDF collaboration has estimated the ratio of fragmentation functions through
semi-inclusive $\bar B \to D \ell^- \bar \nu_\ell X$ decays \cite{bib:CDF_fdfs3}.  
The reconstructed $D \ell^-$ signal yields are then related to the number of 
produced $b$ hadrons by assuming the $SU(3)$ flavor symmetry and neglecting
$SU(3)$-breaking corrections  
(e.g. assuming \ $\Gamma(\bar{B}_d^0\to \ell^-\bar{\nu}_\ell D^+)=\Gamma(\bar{B}^0_s\to \ell^-\bar{\nu}_\ell D_s^+)$). 
Together with an earlier result using double semileptonic decays (containing
two muons and either a $K^*$ or a $\phi$-meson)~\cite{bib:CDF_fdfs1} 
the average value $f_s/(f_d+f_u) = 0.142 \pm 0.019$ is obtained~\cite{PDG}.

An alternative approach uses the different mixing probabilities
for $B^0_d$ and $B^0_s$ mesons. Despite a $1.8 \sigma$ discrepancy in the 
time-integrated mixing probability between the LEP and Tevatron data, an average 
value of $f_s=0.119 \pm 0.019 $ was determined with this method~\cite{bib:HFAG}.

The CLEO and Belle collaborations have extracted
the fraction $f_s$ of $B_s^{(*)}\bar{B}_s^{(*)}$ events among all $b\bar{b}$ events 
at the $\Upsilon(5S)$ resonance 
from inclusive $\Upsilon(5S) \to D_sX, \phi X$ decays~\cite{bib:CLEO_fs,bib:Belle_fs}. 
Here the relation
\begin{eqnarray}
\mbox{BR}(\Upsilon(5S)\to D_sX, \phi X)  = 
2 f_s     \mbox{BR}(B_s^0\to D_sX, \phi X) &&\nonumber \\
+ (1-f_s)\mbox{BR}(\Upsilon(4S) \to D_sX, \phi X) &&
\end{eqnarray}
is assumed with $\mbox{BR}(B_s^0\to D_sX)=(92 \pm 11)\%$, which relies
on a variety of assumptions and yields the model-dependent result 
$f_s = 0.194\pm 0.011 \mbox{(stat)} \pm 0.027 \mbox{(sys)}$~\cite{bib:HFAG}.

It is evident that the fragmentation functions depend on the environment,
which becomes apparent when an attempt is made to compare the 
numerical values for $f_s$. At the $B$ factories $f_d+f_u+f_s=1$, whereas
at hadron colliders the available energy allows the $b$-quark to fragment into baryons
as well.
In addition beam-remnant effects at hadron colliders might affect the
$b$-hadron fractions depending on $p_T$ and/or pseudo-rapidity.
Consequently, each experiment -- and in particular LHCb -- should calibrate its own value 
for this quantity. 
As a result LHCb cannot directly use the value measured at Tevatron or at LEP. 
The fragmentation function is not only the major limiting parameter for the 
determination of $\mbox{BR}(B^0_s\to\mu^+\mu^-)$ at LHCb, but in fact 
for all $B_s$-decay branching ratio measurements at the LHC, the Tevatron, 
and an $e^+e^-$ $B$ factory running at $\Upsilon(5S)$. 
As a result, the usage of $B_s$ modes as normalization channels, obtained from the 
KEKB runs at $\Upsilon(5S)$, also suffer from an imprecise value of $f_s$, in addition to a 
large statistical uncertainty~\cite{bib:Belle_Bs}.

By normalizing the $B_s^{0}\to \mu^+\mu^-$ decay directly to another $B_s$ decay, the
ratio of the fragmentation functions in Eq.~(\ref{BRmumu-exp}) would trivially disappear. 
However, at present 
the best directly measured $B_s$ branching ratio is $\mbox{BR}(B_s^{0}\rightarrow
D_s\pi)=3.67^{+0.35}_{-0.33}(\mbox{stat})^{+0.43}_{-0.42}\pm
0.49(f_s)$~\cite{bib:Belle_Bs}, 
determined with 23.6 fb$^{-1}$ of data at $\Upsilon(5S)$.
Methods are being considered to improve the present knowledge of $f_s$ 
at the $B$ factories~\cite{Sia:2006cq}. However even considering these possible
improvements it is unlikely that they will be sufficient to match the 
required precision of LHCb. A total uncertainty of about $12\%$ could be expected 
for a sample of $120\,\mbox{fb}^{-1}$ (corresponding to the total available statistics) 
and assuming these additional improvements in the determination of $f_s$~\cite{bib:Louvot}. 
Moreover the decay $\bar B_s^{0}\rightarrow D_s^{+}\pi^{-}$ poses experimental
difficulties when used as normalization channel for $B_s^{0}\rightarrow \mu^+ \mu^-$, 
due to the very different decay topology (hadronic final state, number of tracks, flight distance
of the $D_s$, etc.).
A sizable contribution to the uncertainty in the branching ratio estimation due to 
the ratio of the efficiencies in Eq.~(\ref{BRmumu-exp}) must thus be considered. 
An alternative $B_s$ decay channel for the direct normalization would be 
$B_s\rightarrow J/\psi \phi$, which is however affected by a statistical error 
twice as large compared to $\bar B_s^{0}\to D_s^+ \pi^-$.  
Assuming the full statistics presently available at the
$B$ factories of 120 fb$^{-1}$, in combination with the possible
improvements in the determination of $f_s$, at best a total relative error
of $15\%$ could be expected for this decay. 

\begin{figure}[!b]
    \centering
    \begin{tabular}{cc}
      \includegraphics[width=1.6in]{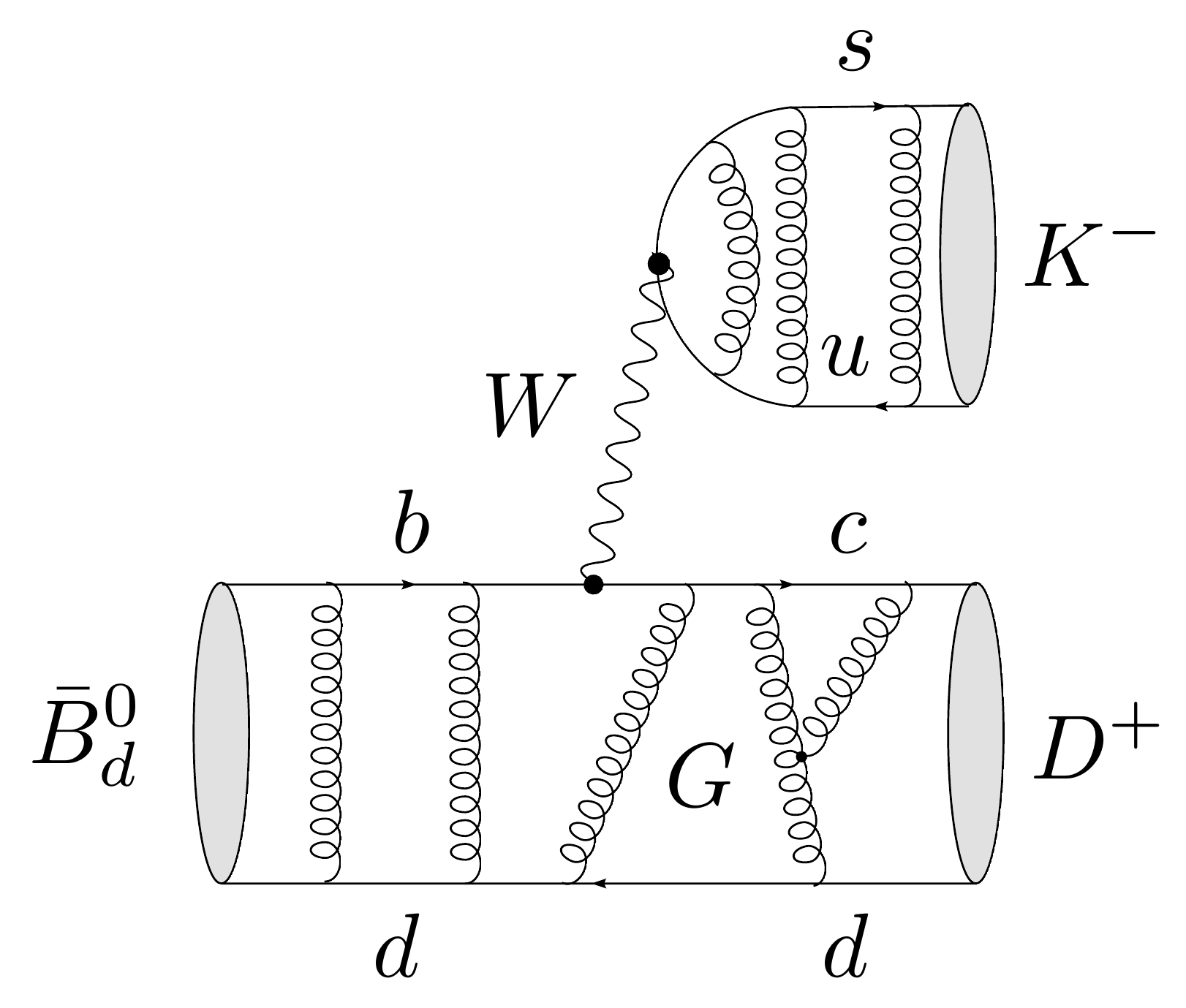} &
      ~ \includegraphics[width=1.6in]{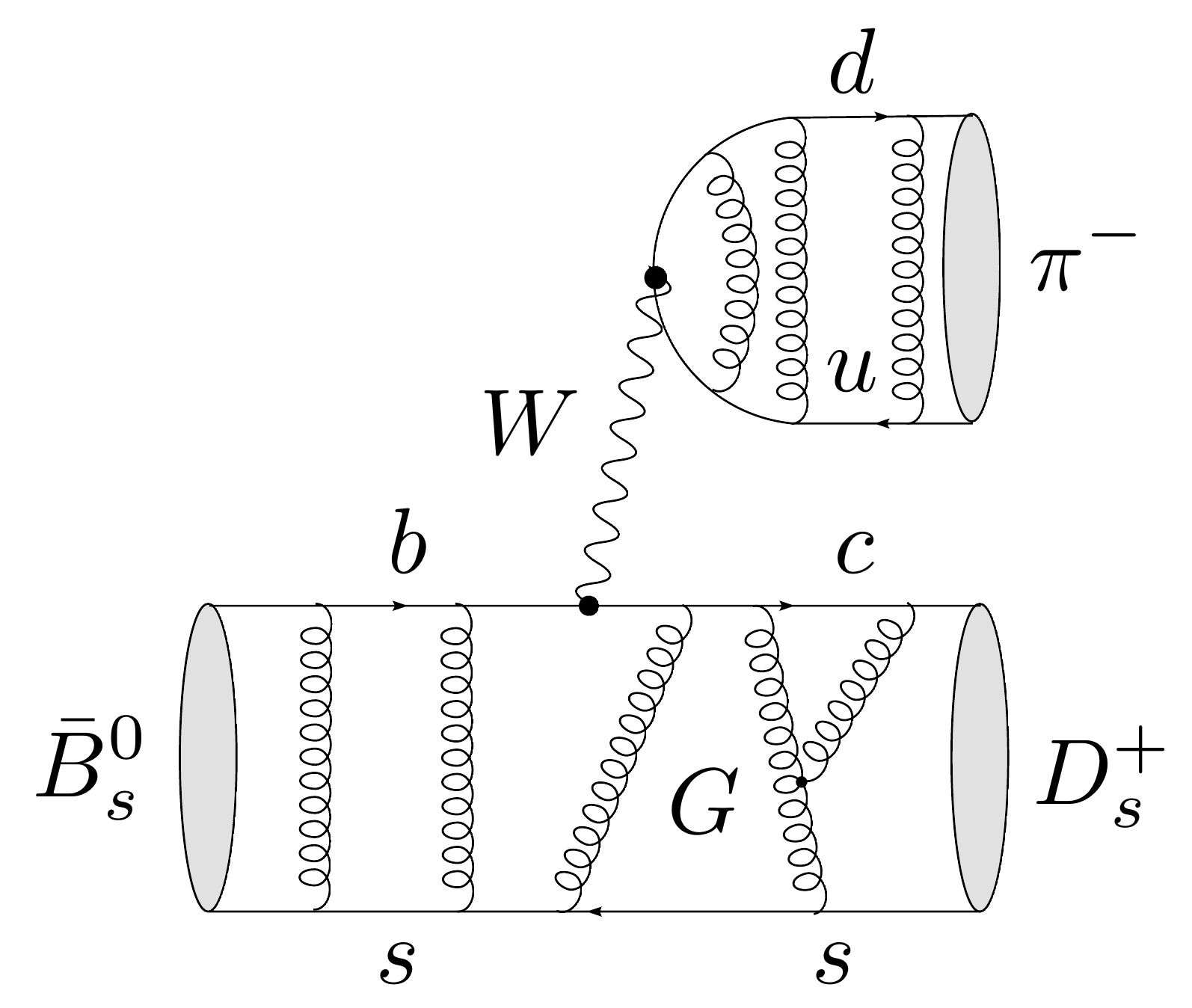}
   \end{tabular} 
    \caption{The $\bar B^0_d\to D^+K^-$ and $\bar B^0_s\to D_s^+\pi^-$ decay
      topologies.}\label{fig:1}
 \end{figure}

%
%
%
\section{A New Strategy for LHCb}\label{sec:strat}
In view of the unsatisfactory situation described in the previous section, we propose a new method for 
extracting $f_d/f_s$ at LHCb. The starting point is the following simple 
expression:
\begin{equation}\label{eq:simple}
\frac{N_s}{N_d} = \frac{f_s}{f_d}\times \frac{\epsilon(B_s\rightarrow
  X_{1})}{\epsilon(B_d\rightarrow X_{2})}\times \frac{\mbox{BR}(B_s\rightarrow 
  X_{1})}{\mbox{BR}(B_d\rightarrow X_{2})};
\end{equation}
knowing the ratio of the branching ratios, we could obviously extract $f_d/f_s$ 
experimentally. In order to implement this feature in practice, the $B_s\to X_1$ and $B_s\to X_2$
decays have to satisfy the following three requirements:
\begin{itemize}
\item[(1)] the ratio of their branching ratios must be easy to measure at LHCb;
\item[(2)]the decays must be robust with respect to the impact of NP contributions;
\item[(3)]the ratio of their branching ratios must be theoretically well understood  within the SM.
\end{itemize}
At first sight, an obvious choice seems to use semileptonic decays such 
as $\bar B \rightarrow D^+ \mu^- \nu$. However, the measurement of such channels at hadron 
colliders is experimentally challenging since the fully reconstructed $B$-mass
is not available and various sources of muons in the background have to be 
controlled. Therefore, we have to focus at non-leptonic decays, where requirement (1) implies to 
look at decays into charged particles and requirement (2) narrows down the search to channels 
without penguin contributions, which are flavor-changing neutral-current processes which
might well be affected by NP contributions. The third requirement finally guides us 
to the decays $\bar B_s^0\rightarrow D_s^+ \pi^-$ and $\bar B_d^0 \rightarrow D^+ K ^-$.

As can be seen in Fig.~\ref{fig:1}, these channels receive only contributions from color-allowed
tree-diagram-like topologies and are related to each other through the interchange of all 
down and strange quarks, i.e.\ through the $U$-spin subgroup of the $SU(3)$ flavor symmetry.
Moreover, the concept of ``factorization'' \cite{fact} is expected to work well 
in these transitions.  This was expected from ``color transparency'' already 
two decades ago \cite{bjor,DG}, while this feature could actually be put on 
a rigorous theoretical basis in the heavy-quark limit \cite{BBNS,SCET}. Consequently, using these
decays, we can calculate the corresponding ratio of their branching ratios entering 
Eq.~(\ref{eq:simple}) up to small, nonfactorizable, $U$-spin-breaking corrections.
This feature will be discussed in more detail in Section~\ref{sec:theo}.

Let us note that in contrast to the $\bar B^0\to D^+\pi^-$ mode usually considered in the 
literature in the context with factorization, the decays in Fig.~\ref{fig:1} have the advantage 
of not receiving additional contributions from ``exchange'' topologies, which are expected 
to be small but are not factorizable. Moreover, thanks to the absence of ``penguin'' topologies, the 
situation concerning factorization is also much more favourable than in $B\to\pi\pi,\pi K$ decays.

Applying a notation similar to that of Ref.~\cite{BBNS},  we write the branching ratios
of the decays at hand as
\begin{eqnarray}
\lefteqn{\hspace*{-0.3truecm}\mbox{BR}(\bar B^0_q \to D_q^+P^-)=
\frac{G_{\rm F}^2(m_{B_q}^2-m_{D_q}^2)^2|\vec q|
\tau_{B_q}}{16\pi m_{B_q}^2}}\nonumber\\
&&\hspace*{-0.3truecm}\times |V_{q}^\ast V_{cb}|^2 
\left[f_P F_0^{(q)}(m_P^2)\right]^2 |a_1(D_qP)|^2\label{BR-th}
\end{eqnarray}
with $P=K$ and $\pi$ for $q=d$ and $s$, respectively. Here $G_{\rm F}$ is
Fermi's constant, the $m$ factors denote meson masses,
$\vec q$
is the momentum of the final-state $D_q$ and $P$ mesons in the rest frame
of the $\bar B^0_q$ meson, $\tau_{B_q}$ is the lifetime of the $\bar B^0_q$,
$V_q^\ast V_{cb}$ with $V_q=V_{us}$ and $V_{ud}$ for $q=d$ and $s$, respectively,
contains the relevant elements of the Cabibbo--Kobayashi--Maskawa (CKM) matrix, 
$f_P$ is $P$-meson decay constant, and the form factor $F_0^{(q)}$ enters
the parametrization of the $\langle D_q^+|\bar c\gamma^\mu b|\bar B^0_q\rangle$
matrix element. The quantity $a_1(D_qP)$ describes the deviation from naive factorization. 
As discussed in detail in Ref.~\cite{BBNS}, this
parameter is found in ``QCD factorization" (QCDF)
as a quasi-universal quantity $|a_1|\simeq 1.05$ with very small 
process-dependent nonfactorizable corrections. 

We would like to propose to measure the ratio of the $\bar B^0_s\to  D_s^+\pi^-$ 
and $\bar B^0_d\to D^+K^-$ branching ratios to determine $f_d/f_s$. Neglecting, 
for simplicity, kinematical mass factors, we have
\begin{eqnarray}
\lefteqn{\frac{\mbox{BR}(\bar B^0_s\to  D_s^+\pi^-)}{\mbox{BR}
(\bar B^0_d\to D^+K^-)}\sim \frac{\tau_{B_s}}{\tau_{B_d}}
\left|\frac{V_{ud}}{V_{us}}\right|^2 }\nonumber\\
&&\times
\left(\frac{f_\pi}{f_K}\right)^2\left[\frac{F_0^{(s)}(m_\pi^2)}{F_0^{(d)}(m_K^2)}
\right]^2\left|\frac{a_1(D_s\pi)}{a_1(D_dK)}\right|^2.\label{eq-rat}
\end{eqnarray}
On the other hand, the ratio of the corresponding number of signal events observed 
in the experiment is given by
\begin{equation}
\frac{N_{D_s\pi}}{N_{D_d K}}=\frac{f_s}{f_d} 
\frac{\epsilon_{D_s\pi}}{\epsilon_{D_d K}}
\frac{\mbox{BR}(\bar B^0_s \to D_s^+\pi^-)}{\mbox{BR}(\bar B^0_d \to D^+K^-)},
\end{equation}
where the $\epsilon$ are again total detector efficiencies.
Using (\ref{BR-th}), we hence obtain
\begin{equation}\label{fs-det}
\frac{f_d}{f_s}=12.88\times\frac{\tau_{B_s}}{\tau_{B_d}}\times
\left[{\cal N}_a {\cal N}_F
\frac{\epsilon_{D_s \pi}}{\epsilon_{D_d K}}
\frac{N_{D_d K}} {N_{D_s\pi}}\right],
\end{equation} 
with
\begin{equation}\label{NF_definition}
 {\cal N}_a \equiv \left|\frac{a_1(D_s\pi)} {a_1(D_dK)}\right|^2,
 \quad {\cal N}_F \equiv \left[\frac{F_0^{(s)}(m_\pi^2)}{F_0^{(d)}(m_K^2)}\right]^2.
\end{equation}
Let us next first explore the experimental feasibility at LHCb before having a closer look at the
theoretical limitations of our new strategy for extracting $f_d/f_s$.

\section{Experimental Prospects at LHCb}\label{sec:exp}
At LHCb, both the $\bar B^0_s\to D_s^+\pi^-$ and $\bar B^0_d\to D^+K^-$ decay
channels can be exclusively reconstructed using the $D^+\to
K^-\pi^+\pi^+$ and $D_s^+\to K^+K^-\pi^+$ final states. An expected
$B$-mass resolution of 18~MeV and excellent particle identification capabilities
will allow LHCb to select and reconstruct a clean sample of these decays.  Since
both channels are selected with an identical flavor final state containing the four
charged hadrons $KK\pi\pi$, the uncertainty on
$\epsilon_{D_s \pi}/\epsilon_{D_d K}$ is expected to be small.

We estimated the corresponding statistical uncertainty on 
$r\equiv \epsilon_{D_s \pi} N_{D_d K}/(\epsilon_{D_d K} N_{D_s\pi})$ 
with a toy Monte Carlo, generating a sample equivalent to $0.2$~fb$^{-1}$.
This is the expected integrated luminosity at the end of 2010,
taking a lower $b\bar{b}$ cross section of 250 $\mu$b due to the reduced LHC
beam energy of 3.5~TeV into account. Following the estimates 
from full simulation~\cite{bib:LHCb-BDh-note}, 
and assuming a total trigger efficiency of 30\%~\cite{bib:LHCbReopTDR},
we expect to
select 5500 $\bar B^0_s\to D_s^+\pi^-$ and 1100 $\bar B^0_d\to D^+K^-$
events, with a background of approximately 6600 $\bar B^0_d\to D^+\pi^-$ events,
where one of the three pions is misidentified as a kaon (assuming a 5\% 
probability to mis-identify a pion as a kaon). Combinatorial
background from inclusive $b\bar{b}$ events is expected to yield 6000 events
inside a mass window $5220 < m < 5420$~MeV around the $B$-mass.  We
expect a precision of $7.5\%$ on $r$, where the dominant uncertainty
originates from $\mbox{BR}(D_s\to K^+K^-\pi)= (5.50 \pm 0.28)\%$.
With an integrated luminosity of 1~fb$^{-1}$ as expected at the end of
2011, the statistical uncertainty becomes negligible, thereby reducing the
total uncertainty to $\sim5.6\%$. 

The ratio $f_d/f_s$ is not only crucial for the precise determination of 
$\mbox{BR}(B^0_s\to\mu^+\mu^-)$ but actually for the measurement of any
$B_s$ branching ratio. Similarly, the general purpose LHC experiments ATLAS 
and CMS rely on a precise value of  $f_d/f_s$ for the determination of 
$\mbox{BR}(B^0_s\to\mu^+\mu^-)$. Unfortunately our proposed hadronic decays 
are not ideal for these experiments due to trigger and particle identification 
requirements.
However, we advocate to apply the value of $f_d/f_s$ as determined by LHCb
also at ATLAS and CMS, once the dependence of $f_d/f_s$ on $p_T$ and/or rapidity 
is measured to be small.

\section{Theoretical Limitations}\label{sec:theo}
In the extraction of $f_d/f_s$ through (\ref{fs-det}), we have theoretical
uncertainties related to $U$-spin-breaking effects in ${\cal N}_a$ and 
${\cal N}_F$. In the case of the first factor, we can write
\begin{equation}\label{Na}
 {\cal N}_a\approx1+2\Re(a_1^{\rm NF}(D_s\pi)-a_1^{\rm NF}(D_dK)),
\end{equation}
where the $a_1^{\rm NF}$ describe the nonuniversal, i.e.\ process-dependent,
nonfactorizable contributions to the decays at hand. These contributions cannot
be calculated reliably. However, they arise as power corrections to the heavy-quark 
limit, i.e.\ they are suppressed by at least one power of $\Lambda_{\rm QCD}/m_b$,  
and are -- in the decays at hand -- numerically expected at the few percent 
level \cite{BBNS}. Let us note that within QCDF, also final-state interaction effects such 
as $\bar B^0_d \to  [D^0 \bar K^0] \to D^+K^-$ arise only as nonfactorizable 
$\Lambda_{\rm QCD}/m_b$ corrections. 

The nonfactorizable terms can actually be probed \cite{bjor} 
through the differential rate of the semileptonic decay 
$\bar B^0_q\to D_q^+\ell^-\bar\nu_\ell$
which yields the following expression  \cite{BBNS}:
\begin{eqnarray}\label{NaEquation}
\lefteqn{\frac{\mbox{BR}(\bar B^0_q \to 
D_q^+P^-)\tau_{B_q}}{d\Gamma(\bar B^0_q\to D_q^+\ell^-\bar\nu_\ell)
/dq^2|_{q^2=m_P^2}}}\nonumber\\
&&=6\pi^2|V_q|^2f_P^2|a_1(D_qP)|^2X_P,
\end{eqnarray}
where $X_P$ deviates from 1 below the percent level. Replacing the pseudoscalar
mesons $P$ by their vector-meson counterparts, i.e.\ $K^- \to K^{*-}$ 
and $\pi^-\to\rho^-$, the corresponding $X_V$ would be exactly given by 1. 
However, these modes are more challenging for LHCb.
The current experimental value
$\mbox{BR}(\bar B^0_d\to D^+K^-)=(2.0\pm0.6)\times10^{-4}$ \cite{PDG} 
agrees well with the number in Ref.~\cite{BBNS}, although the uncertainty
is still too large to probe the nonfactorizable effects. This will be feasible
at LHCb by combining the measurement of the $\bar B^0_d\to D^+K^-$
branching ratio described above with measurements of the differential semileptonic
$\bar B^0\to D^+\ell^-\bar\nu_\ell$ rate at $q^2=M_K^2$ by the $B$-factory 
experiments BaBar and Belle. 

It is interesting to note that factorization was already tested in a similar setting 
at the $B$ factories, measuring the branching ratios and the $D_{(s)}^{*-}$ 
polarization in the decays $B_s^0\to D_s^{*-}\rho^+$ and $B^0\to D^{*-}\omega \pi^+$. 
Good agreement was found between factorization predictions and the experimental 
results within the current errors~\cite{FactBabar,FactBelle}. 

The deviation of (\ref{Na}) from 1 is actually not only suppressed by $\Lambda_{\rm QCD}/m_b$ 
but also through the feature that this is a $U$-spin-breaking difference. In this context, it should 
again be emphasized that any decay topology contributing to $\bar B^0_d\to D^+K^-$, even the 
most complicated rescattering topology, has a counterpart in $\bar B^0_s\to D_s^+\pi^-$,
which is related to the $B_d$ case through the interchange of all down and strange quarks.
Consequently, taking all these considerations into account, we eventually conclude that 
$1-{\cal N}_a$  is conservatively  expected to be at most a few percent.

The major uncertainty affecting (\ref{fs-det}) is hence the 
form-factor ratio ${\cal N}_F$, where $U$-spin-breaking corrections 
arise from $d$ and $s$ spectator-quark effects, which were 
neglected in previous determinations of $f_s$  \cite{bib:CDF_fdfs3}. 
Making the same approximation, we would simply have ${\cal N}_F=1$.
Unfortunately, the $B_s\to D_s$ form factors have so far received only small 
theoretical attention. 
In Ref.~\cite{chir}, such effects were explored using heavy-meson chiral 
perturbation theory, while QCD sum-rule techniques were applied in 
Ref.~\cite{BCN}. The numerical value given in the latter paper yields 
${\cal N}_F=1.3\pm0.1$.

Interestingly, we can obtain a {\it lower} bound on the
$B^0_s\to\mu^+\mu^-$ branching ratio from our strategy. 
Using (\ref{BRmumu-exp}) and (\ref{fs-det}) and assuming ${\cal N}_a=1$ yields
\begin{equation}
\mbox{BR}(B^0_s\to\mu^+\mu^-)={\cal N}_{F}
\mbox{BR}(B^0_s\to\mu^+\mu^-)_0,
\end{equation}
where $\mbox{BR}(B^0_s\to\mu^+\mu^-)_0$ follows from the analysis
described above by assuming vanishing $U$-spin-breaking corrections. Since the 
radius of the $B^0_s$ is smaller than that of the $B^0_d$, we expect 
${\cal N}_{F}>1$ \cite{chir}. This behaviour is actually reproduced in the 
calculation of the chiral logarithms in Ref.~\cite{chir}, as well as in the 
QCD sum-rule calculation in Ref.~\cite{BCN}. Moreover, the sign of the 
chiral logarithmic correction to the $SU(3)$-breaking ratio of the decay
constants of $D_{(s)}$ and $B_{(s)}$ mesons agrees with experimental 
(for $D_{(s)}$) and lattice results (and also the numerical values are found
of similar size). The inequality ${\cal N}_{F}>1$ implies then the following 
bound:
\begin{equation}\label{BR-bound}
\mbox{BR}(B^0_s\to\mu^+\mu^-)>\mbox{BR}(B^0_s\to\mu^+\mu^-)_0,
\end{equation}
which offers an interesting tool for the detection of possible NP contributions to
$B^0_s\to\mu^+\mu^-$ at LHCb. Assuming that we will measure 
$\mbox{BR}(B^0_s\to\mu^+\mu^-)_0$ to be $5\sigma$ above the
SM prediction (\ref{Bsmumu-SM}), $U$-spin-breaking effects could only enhance
the measured branching ratio and could {\it not} move it down towards the SM
value. 

In the long run, we would of course like to measure $\mbox{BR}(B^0_s\to\mu^+\mu^-)$
as accurately as possible. In order to match the experimental precision for $r$
of about 5\% discussed above, it is sufficient to know the $U$-spin-breaking 
corrections to the form-factor ratio $F_0^{(s)}(m_\pi^2)/F_0^{(d)}(m_K^2)$ 
from nonperturbative calculations, such as lattice QCD, at the level of 20\%. 
This looks feasible to us, in particular in view of the tremendous amount of work 
that was invested to study $B\to D$ form factors on the lattice for the extraction 
of $|V_{cb}|$ from semi-leptonic $B\to D \ell\bar\nu_\ell$ decays. We are not aware 
of any lattice calculation of the $SU(3)$-breaking corrections to the form-factor
ratio entering our strategy, which is due to the fact that such analyses did so far 
not appear phenomenologically interesting. 

Finally, we would like to note that the $SU(3)$-breaking effects in Eq.~(\ref{eq-rat}) coming from
the ratios of decay constants and form factors tend to cancel each other. Assuming
${\cal N}_F=1.3$ with $f_\pi/f_K=0.8$, we get an overall $SU(3)$-breaking correction to
the ratio of branching ratios of only about 10\%, which is surprisingly small and suggests that also
the $SU(3)$ suppression of ${1-\cal N}_a$ is very efficient. 

\begin{figure}[!t]
  \centering
  \begin{tabular}{cc}
    \includegraphics[width=0.245\textwidth]{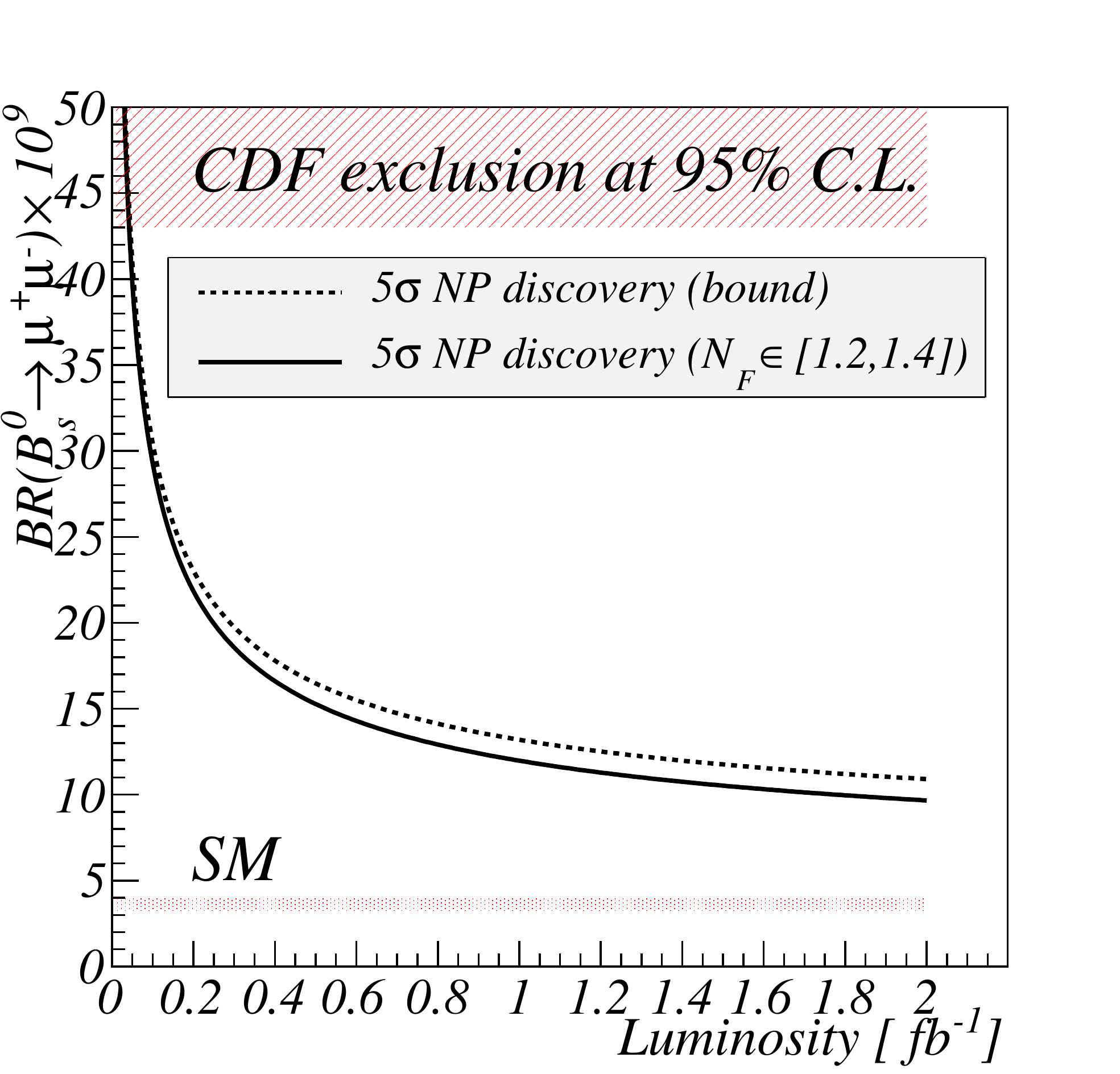} 
    \includegraphics[width=0.245\textwidth]{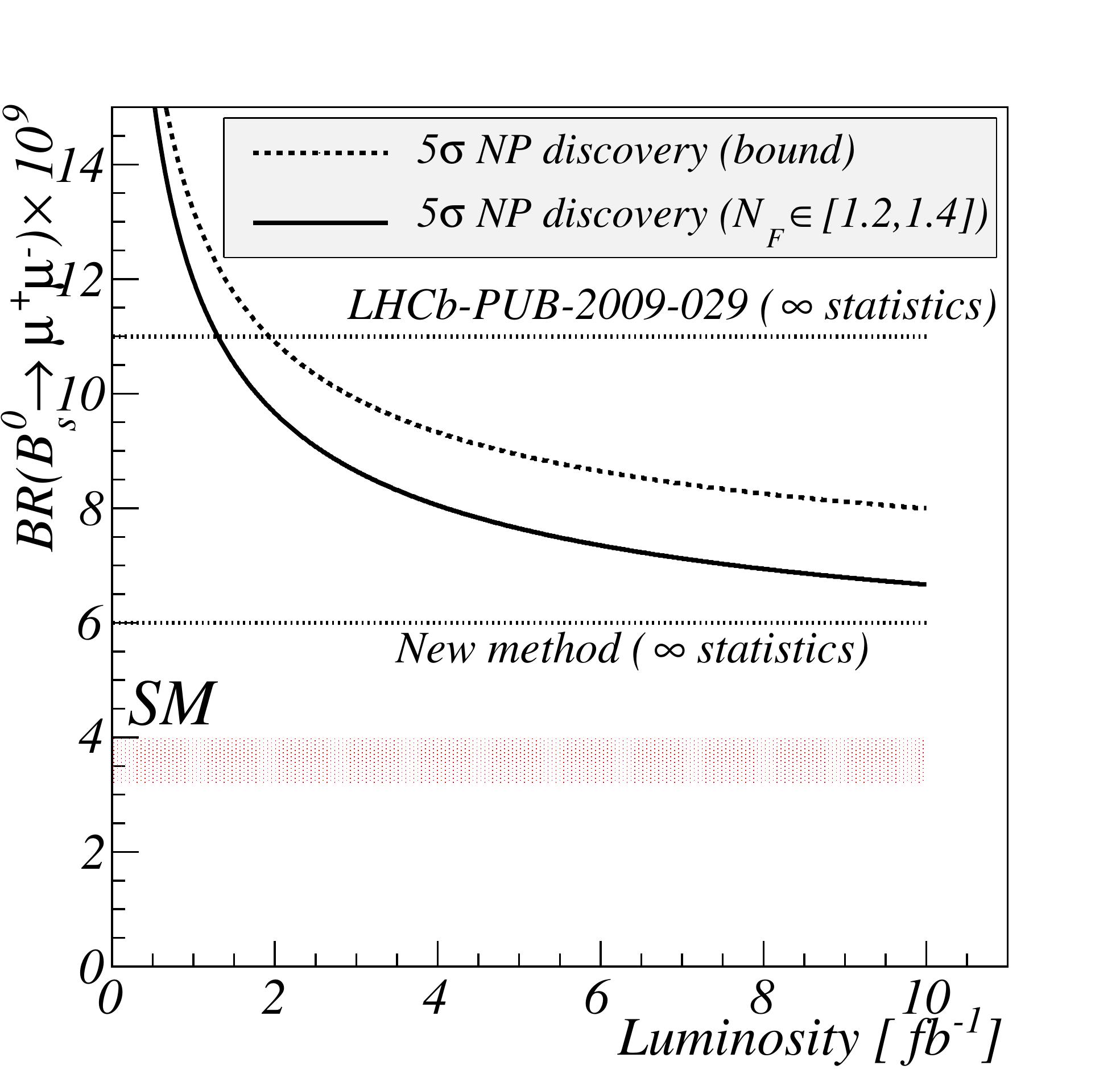} 
  \end{tabular} 
  \caption{(color online). Illustration of the LHCb NP discovery potential in $B^0_s\to\mu^+\mu^-$
  resulting from our strategy (${\cal N}_F\in [1.2,1.4]$). We show the smallest value of
  $\mbox{BR}(B^0_s\to\mu^+\mu^-)$ allowing the detection of a $5 \sigma$ deviation
  from the SM as a function of the luminosity at LHCb (at the nominal beam 
  energy of 14 TeV). The figure on the left-hand side shows the low-luminosity 
  regime, whereas the one on the right-hand side illustrates the asymptotic 
  behaviour (curves extrapolated from Ref.~\cite{bib:LHCbRoadMap}).}\label{fig:2}
\end{figure}

\section{Implications for the NP Reach in the Measurement of
$\mbox{BR}(B_s\to\mu^+\mu^-)$}\label{sec:impl}

In Fig.~\ref{fig:2}, we illustrate the NP discovery potential in $B^0_s\to\mu^+\mu^-$ 
at LHCb through our method. We show contours corresponding to a $5 \sigma$ 
NP signal with respect to (\ref{Bsmumu-SM}) for the bound in (\ref{BR-bound}) 
and the extracted value of the branching ratio. Here we have assumed 
that the uncertainty on $\mbox{BR}(D_s\to K^+K^-\pi)$ in the determination
of  $f_d/f_s$ is distributed Gaussian, and likewise for the uncertainty on
$\mbox{BR}(B^0_d\to J/\psi K^*)$ in the extraction of 
$\mbox{BR}(B^0_s\to\mu^+\mu^-)$.
We conservatively varied ${\cal N}_F\in [1.2,1.4]$, resulting in a negligible 
change to the predicted sensitivity. Similarly, a variation of 
${\cal N}_a\in [0.97,1.03]$ does essentially not affect the contour.

As can be seen in the plot on the right-hand side in Fig.~\ref{fig:2},
the resulting NP discovery potential is about
twice as large as the present LHCb expectation~\cite{bib:LHCbRoadMap}
(upper horizontal line) enabling a possible
discovery of NP down to $\mbox{BR}(B^0_s\to\mu^+\mu^-) > 6\times 10^{-9}$ 
(lower horizontal line).
In addition to the increased sensitivity in the regime of low branching ratios, 
even for large values close to the current CDF exclusion limit the significance 
of a possible NP discovery would be increased. Thanks to the decrease of the 
systematical uncertainty, LHCb will be able to fully exploit the statistical 
improvement, taking full advantage of the accumulated LHCb 
data up to 10~fb$^{-1}$, which corresponds to five years of nominal LHCb data taking.
 
At a future LHCb upgrade, even the $B^0_d\to\mu^+\mu^-$ decay will become 
accessible, and the proposed determination of $f_d/f_s$ will be an important tool for the
measurement of $\mbox{BR}(B^0_d\to\mu^+\mu^-)/\mbox{BR}(B^0_s\to\mu^+\mu^-)$,
which provides an even stronger test of the SM~\cite{buras}.

Let us finally emphasize that a future Super-$B$ factory running at $\Upsilon(5S)$ 
would allow us to check the calculations of the $SU(3)$-breaking effects in the 
form factor through the measurement of $\bar B^0_s \to D_s^+ \ell \bar\nu_\ell$ 
decays. The possible discovery of NP in $B^0_s\to\mu^+\mu^-$ at LHCb
does not rely on this input, but constraining -- and even extracting -- 
$SU(3)$-breaking form-factor ratios would lead to a 
more precise determination of $\mbox{BR}(B^0_s\to\mu^+\mu^-)$.

\vspace*{0.6truecm}

\section{Conclusions}\label{sec:concl}
The current experimental knowledge of the ratio $f_d/f_s$ of fragmentation functions 
is unsatisfactory and affects all absolute $B_s$-decay branching ratio measurements at 
hadron colliders. In particular, this quantity is also the major uncertainty for the extraction of the 
$B^0_s\to\mu^+\mu^-$ branching ratio from the LHCb data. In view of this situation, 
we have proposed a new strategy for determining $f_d/f_s$ at LHCb. It uses the pair
of the color-allowed, $U$-spin-related tree decays $\bar B^0_s\to D_s^+\pi^-$ and 
$\bar B^0_d\to D^+K^-$, which are very favorable from an experimental point of view
for LHCb, robust with respect to NP contributions and theoretically well understood,
thereby offering a precise measurement of $f_d/f_s$ at LHCb. The resulting decrease 
of the total systematic uncertainty on $f_d/f_s$ allows us to detect a $5 \sigma$ NP 
signal in the measured $B^0_s\to\mu^+\mu^-$ branching ratio for values as small as t
wice the SM value. This corresponds to an improvement of the 
corresponding NP reach by a factor of 2 with respect to the present LHCb expectation.
Once the dependence of $f_d/f_s$ on $p_T$ and/or rapidity is measured to be small,
the value of $f_d/f_s$ as determined at LHCb by means of our strategy can also be 
applied at ATLAS and CMS.

\section*{Acknowledgements} 
We would like to thank Wouter Hulsbergen, Marcel Merk and Gerhard Raven 
for valuable discussions and carefully reading the manuscript, and Barbara Storaci 
and Remi Louvot for many useful conversations. 
This work is supported by the Netherlands Organisation for Scientific Research (NWO).

\end{document}